# Anisotropic response of defect bound states to magnetic field in epitaxial FeSn films

Huimin Zhang[1,2,3*], Zhengfei Wang[4], Michael Weinert[5], and Lian Li[1*]

[1]Department of Physics and Astronomy, West Virginia University, Morgantown, WV 26506, USA

[2]State Key Laboratory of Structural Analysis, Optimization and CAE Software for Industrial Equipment, Dalian University of Technology, Dalian 116024, China

[3]Department of Physics, Dalian University of Technology, Ministry of Education, Dalian 116024, China

[4]Hefei National Research Center for Physical Sciences at the Microscale, CAS Key Laboratory of Strongly-Coupled Quantum Matter Physics, Department of Physics, Hefei National Laboratory, University of Science and Technology of China, Hefei, Anhui 230026, China

[5]Department of Physics, University of Wisconsin, Milwaukee, WI 53201, USA

[*]Correspondence to: huiminzhang@dlut.edu.cn, lian.li@mail.wvu.edu

**Abstract**

Crystal defects, whether intrinsic or engineered, drive many fundamental phenomena and novel functionalities of quantum materials. Here, we report symmetry-breaking phenomena induced by Sn-vacancy defects on the surface of epitaxial Kagome antiferromagnet FeSn films using low-temperature scanning tunneling microscopy and spectroscopy. Near the Sn-vacancy defects, anisotropic quasiparticle interference patterns are observed in the differential conductance dI/dV maps, indicating two-fold electronic states that break the 6-fold rotational symmetry of the Kagome layer.  Furthermore, the Sn-vacancy defects induce bound states that exhibit anomalous Zeeman shifts under an out-of-plane magnetic field, where their energy shifts linearly towards higher energy independent of the direction of the magnetic field.  Under an in-plane magnetic field, the shift of the bound state energy also shows a two-fold oscillating behavior as a function of the azimuth angle. These findings demonstrate defect-enabled new functionalities in Kagome antiferromagnets for potential applications in nanoscale spintronic devices.

**Introduction**

Defects in crystals are ubiquitous, whether they form during crystal growth or are induced by post-growth processing. These defects can significantly modify the physical and electronic properties of quantum materials, enabling new and tunable functionalities. For example, the nitrogen-vacancy (NV) color centers in Diamond have been identified as a promising candidate for quantum computing qubits due to their optically addressable electron spin states and long coherence time[1–3]. In high-temperature superconductors, nonmagnetic impurities can break up Cooper pairs and introduce bound states within the superconducting gap, which provides insights into the nature of unconventional superconductivity[4–7]. Phase-referenced quasiparticle interference (QPI) technique[7–11] was developed to determine the unconventional pairing symmetry based on the scattering by these nonmagnetic defects.

The recent studies of defect-induced bound states in Kagome materials have also provided significant insight into the interplay of topology, charge-ordered phases, and magnetism. The Kagome lattice, which forms the building block of these materials, is a two-dimensional (2D) network of corner-sharing triangles and hexagons (Fig. 1**a**). This unique structure gives rise to linearly dispersing Dirac cones at the Brillouin Zone (BZ) corner K point, Von Hove singularities at the M point, and a flat band across the entire BZ[12]. These electronic band features have already been confirmed by angle-resolved photoemission spectroscopy (ARPES) in binary metallic Kagome magnets $T_mX_n$ (T: 3d transition metals, X: Sn, Ge, m:n = 3:1, 3:2, 1:1)[13–16], and ternary ferromagnetic $YMn_6Sn_6$[17]. Fascinating phenomena have also been reported, such as the giant spin-orbit tunability of the Dirac mass and electronic nematicity in $Fe_3Sn_2$[18,19], the magnetic Weyl semimetal state and negative flat band magnetism in $Co_3Sn_2S_2$[20–22], the topological Chern magnet in the quantum limit in $TbMn_6Sn_6$[23], and the orbital Zeeman effect in $TbV_6Sn_6$[24]. Furthermore, nonmagnetic substitutional Indium impurity in $Co_3Sn_2S_2$ was shown to introduce spin-polarized bound states consistent with a negative orbital magnetization[25]. Sulfur-vacancy in $Co_3Sn_2S_2$ also exhibits negative orbital magnetization, which was attributed to spin-orbit polarons[26,27].

For antiferromagnetic (AFM) Kagome materials, studies have been focused on FeGe, $Mn_3Sn$, and FeSn. In addition to the *c*-axis collinear AFM ordering below $T_N \approx 410$ K and a double cone (canted)

AFM structure below $T_{Canting} \approx 60$ K[28,29], the FeGe surface exhibits edge states and a (2 × 2) charge order[30], which couples strongly to magnetism based on ARPES and neutron scattering studies[31,32]. On the other hand, Mn$_3$Sn shows an-plane non-collinear AFM order[33,34], leading to large anomalous Hall[35], anomalous Nernst[36], and magneto-optical Kerr[37] effects. Remarkably, all these effects are observed at a moderate magnetic field, making it appealing to control the topological electronic states for AFM-based spintronics. For FeSn, which consists of alternatively stacked planes of the 2D Fe$_3$Sn Kagome (K) layer and Sn$_2$ stanene (S) layer [38,39,40], Fe atoms within each Kagome layer exhibit in-plane ferromagnetic order[38,39], while the neighboring Kagome layers are coupled antiferromagnetically along the *c*-axis with a Néel temperature $T_N \approx 366$ K. For cleaved bulk materials, previous ARPES studies have revealed linearly dispersed Dirac crossings at –0.43 eV and –0.31 eV at K points, and flat bands at –0.23 eV[14]. Recently, we have reported symmetry-breaking electronic nematic order tunable by an applied magnetic field[41] and giant periodic pseudo-magnetic fields greater than 1000 T in FeSn epitaxial films[42].

In this study, we show anisotropic QPI patterns near the single Sn vacancy defects in the differential conductance dI/dV maps on the K-terminated epitaxial FeSn films, which breaks the 6-fold rotational symmetry of the Kagome layer. Furthermore, Sn-vacancy defects induce bound states that exhibit anomalous Zeeman shifts under a vertical magnetic field, where their energy shifts linearly towards higher energy independent of the direction of the magnetic field. Under an in-plane magnetic field, the shift of the bound state energy also shows a two-fold oscillating behavior as a function of the azimuth angle, demonstrating the potential for field-controlled anisotropic transport for nanoscale spintronic devices.

**RESULTS AND DISCUSSION**

***Point defects in epitaxial kagome antiferromagnet FeSn films.*** The epitaxial FeSn/STO(111) films exhibit island growth with two kinds of terminations (Fig. 1**a** and Fig. S1), one with a honeycomb lattice in Fig. 1**b** and the other with a close-packed lattice in Fig. 1**c**, assigned to the S- and K-layers, respectively. Point defects are often seen on the S-layer, which typically shows an enhanced local density of states (LODS) with two-fold symmetry, labeled as type I (Figs. 1**b** and S2). On the other hand, defects on the K-layer are often associated with a suppressed density of states, labeled as type II (Figs. 1**c** and S3). To determine the nature of these defects, we carried out first-principles density functional theory (DFT) calculations. The simulated LDOS for the S/K layer with Sn vacancy defects agrees well with the experimental STM images (Fig. S4 and Supplementary Note 1). Thus, the Type I and II defects are attributed to Sn divacancy and Sn single vacancy on the S and K layers, respectively.

***Symmetry-breaking QPI pattern near the Sn vacancy defects.*** The dI/dV spectra further reveal the differences between the two types of defects (Figs. 1**d**&**e**). Type I defect shows a bound state at -89.7 meV in the dI/dV spectrum (red curve in Fig. 1**d**). In comparison, type II defect displays such a state at -19.0 meV (red curve in Fig. 1**e**). We notice that while the bound state energy for the Sn divacancy defect in S layer is at the specific value, for the Sn single vacancy in K layer, the energy is spatially dependent as shown in Figs. 1**f-g** and also Fig. S5, likely due to coupling with neighboring defects similar to earlier studies[27]. Both Sn vacancy defects lead to anisotropic QPI patterns near the defects in the differential conductance dI/dV maps, representing the density of states. For the Sn divacancy on the S-layer, the two-fold symmetrical bound states seen in the topographic images (Fig. S2) and dI/dV maps (Fig. S6) are consistent with the divacancy character of the defect.

In contrast, for the Sn-vacancy on the K layer, the spatial distribution of the bound state displays a complex energy dependence: a trimer feature within the energy ranges from -140 to -40 meV and a dimer feature at the energy of -20 meV (Fig. 2**b**). Interestingly, the bound states become featureless from the Fermi level to 20 meV and a contrast reversal from bright to dark occurs at ~120 meV (more details in Fig. S7). In addition, the defect induces a symmetry-breaking QPI pattern as a triangular-shaped depression in differential conductance maps (c.f., $g(\mathbf{r}, -140\ \text{meV})$). As outlined by the dotted triangles sized 3 × and 5 × lattice constant, QPI patterns flip upside down at 20 meV, suggesting a two-fold electronic state that breaks the 6-fold rotational symmetry of the Kagome layer. This signature of electronic nematicity[41], is further supported by the anisotropic response of the bound states to the magnetic field discussed below.

***Anomalous Zeeman shift under an out-of-plane magnetic field.*** Next, we examine the response of the defect bound states to the out-of-plane magnetic field ($B_\perp$). For Sn divacancy (type I), the dI/dV spectra taken at the same site (red dot in Fig. 3**a**) under various $B_\perp$ is shown in Figs. 3**b&c,** and the peak position is summarized in Fig. 3**d**. The bound states experience a negative energy shift between -0.5 T and 0.5 T (here, the positive direction is defined as $\Delta E = (E_B - E_{B=0}) < 0$), independent of the direction of the magnetic fields. Specifically, the peak position shifts by $\Delta E$ = 4.9 meV away from the Fermi level at $B_\perp$ = -9 or 9 T compared to that under $B_\perp$ = 0 T, as shown in Fig. 3**c**. Before the peak position saturates above $B_\perp$ = 0.5 T, the shift can be well fitted by the linear function $\Delta E_B = g \cdot \Delta B$, which yields a slope of 4.79 ± 0.28 meV·T$^{-1}$, corresponding to an effective $g$ factor of 165.2 ± 9.7 (Fig. 3**d** and Supplementary Note 2). Similar behavior is seen for the single-Sn vacancy defect on the K-layer (Type II) but with a relatively smaller slope 2.36 ± 0.12 meV·T$^{-1}$, corresponding to an effective $g$ factor of 81.4 ± 4.2 (Figs. 3**e-h**). The identical switch directions with the our-of-plane magnetic fields indicate an anomalous Zeeman shift in both S- and K-layers, similar to that previously reported in Kagome magnets $Co_3Sn_2S_2$[22,25,26], $Fe_3Sn_2$[18], and $TbV_6Sn_6$[24]. Nevertheless, they are much larger than these of in $Co_3Sn_2S_2$ (0.075 meV·T$^{-1}$, 0.174 meV·T$^{-1}$) [22,25,26], and smaller than these in $Fe_3Sn_2$ (12 meV·T$^{-1}$)[18] and $TbV_6Sn_6$ (11.20 meV·T$^{-1}$).

For the normal Zeeman effect, $\Delta E = -\mu \cdot B$, where $\Delta E$ is the energy shift, $\mu$ the magnetic moment, and $B$ the applied magnetic field. The energy would decrease if the magnetic moment $\mu$ is parallel to the applied magnetic field $B$ but would increase if antiparallel. However, when the energy shift $\Delta E$ is independent of the magnetic field direction, the so-called anomalous Zeeman shift $\Delta E < 0$ indicates that the net magnetization of the bound states is always parallel to the direction of the magnetic field, suggesting strong contributions from the orbital moment (Fig. S8). Previously, positive energy shift ($\Delta E > 0$) has been observed in magnetic Weyl semimetal $Co_3Sn_2S_2$, either for the flat band feature[22], or defect induced bound states (Indium-doped or Sulfur-vacancy)[25,26]. In contrast, negative energy shift ($\Delta E < 0$) was reported in the flat band of ferromagnetic Kagome metal $Fe_3Sn_2$[22]. The anomalous Zeeman shift also saturates at $B$ = 1 T in FeSn (this work) or $Fe_3Sn_2$[18] and $TbV_6Sn_6$[24], but never does, even at $B$ = 8 T in $Co_3Sn_2S_2$[22,25,26].

***Oscillating behavior under in-plane magnetic fields.*** Furthermore, we measured the shift of the defect bound state energy as a function of the azimuth angle $\psi$ with an in-plane magnetic field $B_{//}$ of 1 T (Figs. 4**a&b**). The peak energy positions of the defect bound states as a function of $\psi$ are in Figs. 4**c**&**d**, which can be well fit by a sin function. The oscillatory behavior shows a two-fold symmetry in both the S- and K-layers, which is even more evident in the angular polar plots in Figs. 4**e**&**f**. The anisotropy directly reflects the breaking of the six-fold symmetry of the Kagome lattice. A rotation angle of 29.2° is seen between the spatial anisotropy on the S- (Fig. 4**e**) and K-layers (Fig. 4**f**), consistent with the 30° rotation between the hexagon units of the S and K layers (upper panels of Figs. 4**a**&**b**).

One possible mechanism for this anisotropic response of defect bound states to magnetic field could be that the defect states exhibit a Stoner-Wohlfarth reorientation due to the external field[43] (see detailed discussion in the Supplementary Note 3). However, this does not explain the anisotropic energy-dependent QPI behavior shown in Fig. 2. A more likely scenario is that in the presence of strong spin-orbit coupling, the charge ordering in Kagome materials is strongly coupled to magnetic field. For example, in ferromagnetic $Fe_3Sn_2$, surface states and QPI patterns have shown a similar two-fold anisotropy under an in-plane magnetic field[18]. In antiferromagnetic FeSn[41] and FeGe[30,31], strong magnetic field tunable stripe order and charge density waves have been reported. Even in non-magnetic $AV_3Sb_5$ (A = K, Rb, Cs) Kagome materials, in-plane field tunable superconductivity, charge density waves, and magneto-optical Kerr effect were also observed. Specifically, thin-flake $RbV_3Sb_5$ exhibits a two-fold symmetric superconductivity with an in-plane magnetic field[44], and the Kagome metal $CsV_3Sb_5$ presents a two-fold rotational symmetrical *c*-axis resistivity in both the superconducting and normal states[45]. A recent magneto-optical Kerr effect measurements of $AV_3Sb_5$ further reveal three-state nematicity, suggesting time-reversal symmetry-breaking[46]. Moreover, in strong-correlated materials such as $Sr_3Ru_2O_7$, a compass-like manipulation of electronic nematicity by in-plane magnetic fields has also been observed[47]. In all cases, strong spin-orbit coupling underpins the entanglement of charge orders and magnetic field, leading to various anisotropic responses to in-plane magnetic field.

**CONCLUSIONS**

In summary, the Sn-vacancy defects induce anisotropic QPI patterns in the differential conductance maps on the Kagome layer in epitaxial FeSn films, indicating two-fold electronic states that break the 6-fold rotational symmetry of the Kagome layer. The Sn-vacancy defects also induce bound states that show anomalous Zeeman shifts under an out-of-plane magnetic field. Under an in-plane magnetic field, the shift of the bound state energy further exhibits a two-fold oscillating behavior as a function of the azimuth angle. These findings demonstrate the feasibility of defect engineering to control electronic states in Kagome antiferromagnets for potential applications in nanoscale spintronic devices.

## METHODS

**Sample preparation.** Our STM/S experiments were conducted in a Unisoku ultrahigh vacuum LT-STM system interconnected to a molecular beam epitaxy (MBE) chamber. The FeSn/SrTO$_3$(111) films were prepared by MBE following our previous recipe. The SrTiO$_3$(STO)(111) substrates (Nb-doped 0.05 wt.%) were first degassed at 600 °C for 3 hrs and followed by annealing at 950 °C for 1 hr. During the MBE growth, high-purity Fe (99.995%) and Sn (99.9999%) were evaporated from Knudson cells on STO(111) substrate with temperatures between 480 and 530 °C.

**LT-STM/S characterization.** All the STM/S results were measured at T = 4.5 K. A polycrystalline PtIr tip was used, which was tested on Ag/Si(111) films before the STM/S measurements. The dI/dV tunneling spectra were acquired using a standard lock-in technique with a small bias modulation $V_{mod}$ at 732 Hz.

## ASSOCIATED CONTENT

Supporting Information

## AUTHOR INFORMATION

**Corresponding Author**

**Huimin Zhang** − Department of Physics and Astronomy, West Virginia University, Morgantown, West Virginia 26506, United States

State Key Laboratory of Structural Analysis, Optimization and CAE Software for Industrial Equipment, Dalian University of Technology, Dalian 116024, China

Department of Physics, Dalian University of Technology, Ministry of Education, Dalian 116024, China

Phone: (+86) 18010090159; E-mail: huiminzhang@dlut.edu.cn

**Lian Li** – Department of Physics and Astronomy, West Virginia University, Morgantown, West Virginia 26506, United States

Phone: (+1) 304-293-4270; E-mail: lian.li@mail.wvu.edu

**Authors**

**Zhengfei Wang** – Hefei National Research Center for Physical Sciences at the Microscale, CAS Key Laboratory of Strongly-Coupled Quantum Matter Physics, Department of Physics, Hefei National Laboratory, University of Science and Technology of China, Hefei, Anhui 230026, China

**Michael Weinert** – Department of Physics, University of Wisconsin, Milwaukee, WI 53201, USA

**Author contributions**

L.L. and H.Z. conceived and organized the study. H.Z. performed the MBE growth and STM/S measurements. All authors analyzed the data, and H.Z. and L.L. wrote the paper.

**Notes**

The authors declare no competing financial interest.

## ACKNOWLEDGMENTS

HZ acknowledges support from the National Key Research and Development Program of China (2023YFB3809600), the National Natural Science Foundation of China (Grant No. 12304210), and the Fundamental Research Funds for the Central Universities (DUT24RC(3)015, DUT22LAB104, DUT22ZD103, DUT24LK007). MW and LL acknowledge support from the U.S. Department of Energy, Office of Basic Energy Sciences, Division of Materials Sciences and Engineering under Award No. DE-SC0017632 and the US National Science Foundation under Grant No. EFMA-1741673.

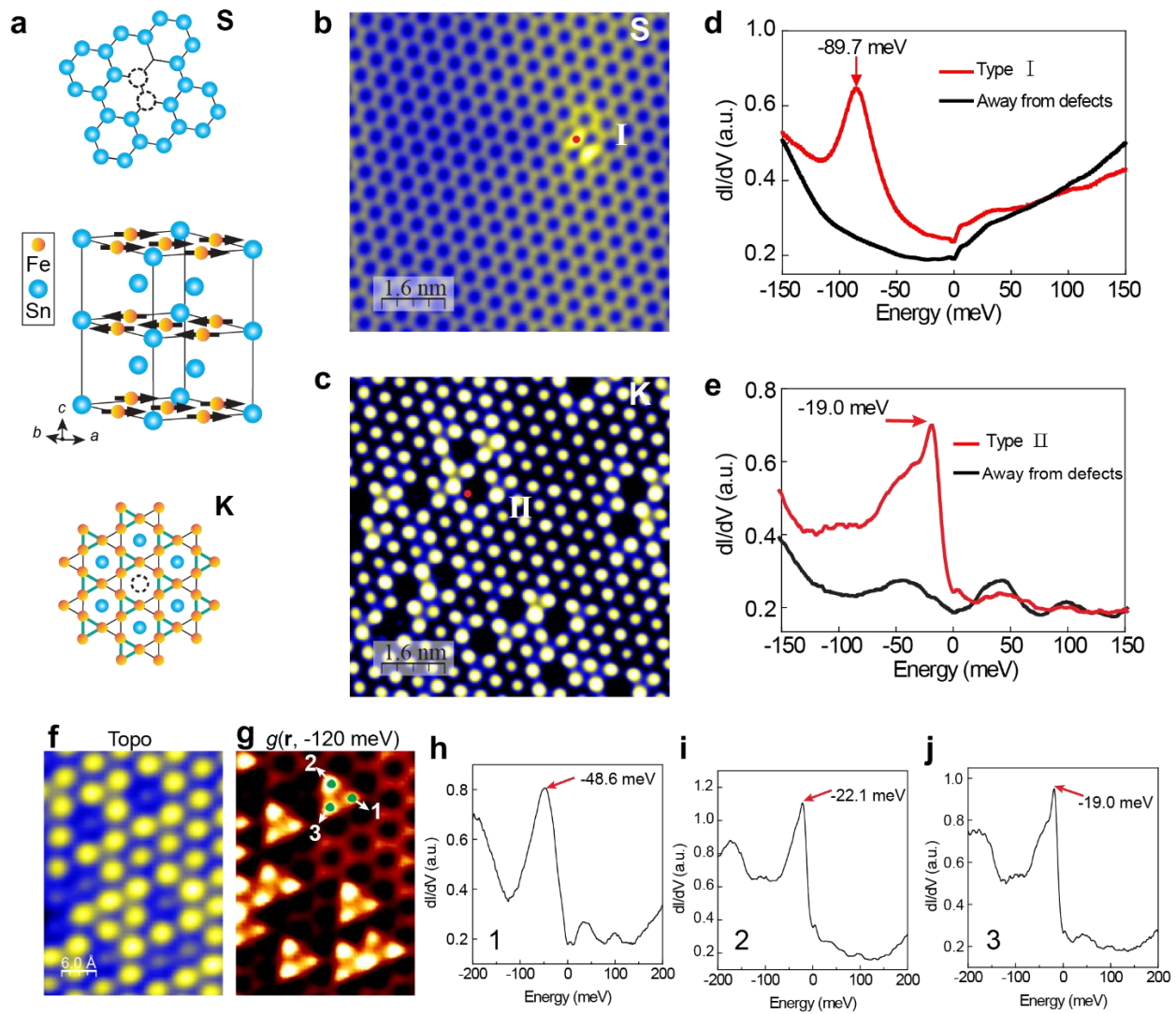

**Figure 1. STM imaging and spectroscopy of MBE-grown FeSn films on the SrTiO$_3$ (111) substrate.**
**a**, The ball-and-stick model of the FeSn, showing the vertical stack of the Stanene (S) and Kagome (K) layers. **b-c**, Topographic STM images of two terminations, S layer (**b**) and K layer (**c**). Setpoint: $V$ = 0.5 V, $I$ = 3.0 nA (**b**) and $V$ = -0.5 V, $I$ = 3.0 nA (**c**). **d-e**, differential conductance dI/dV spectra taken on type I and type II defects (red curve) compared to that taken away from defects (black curve).

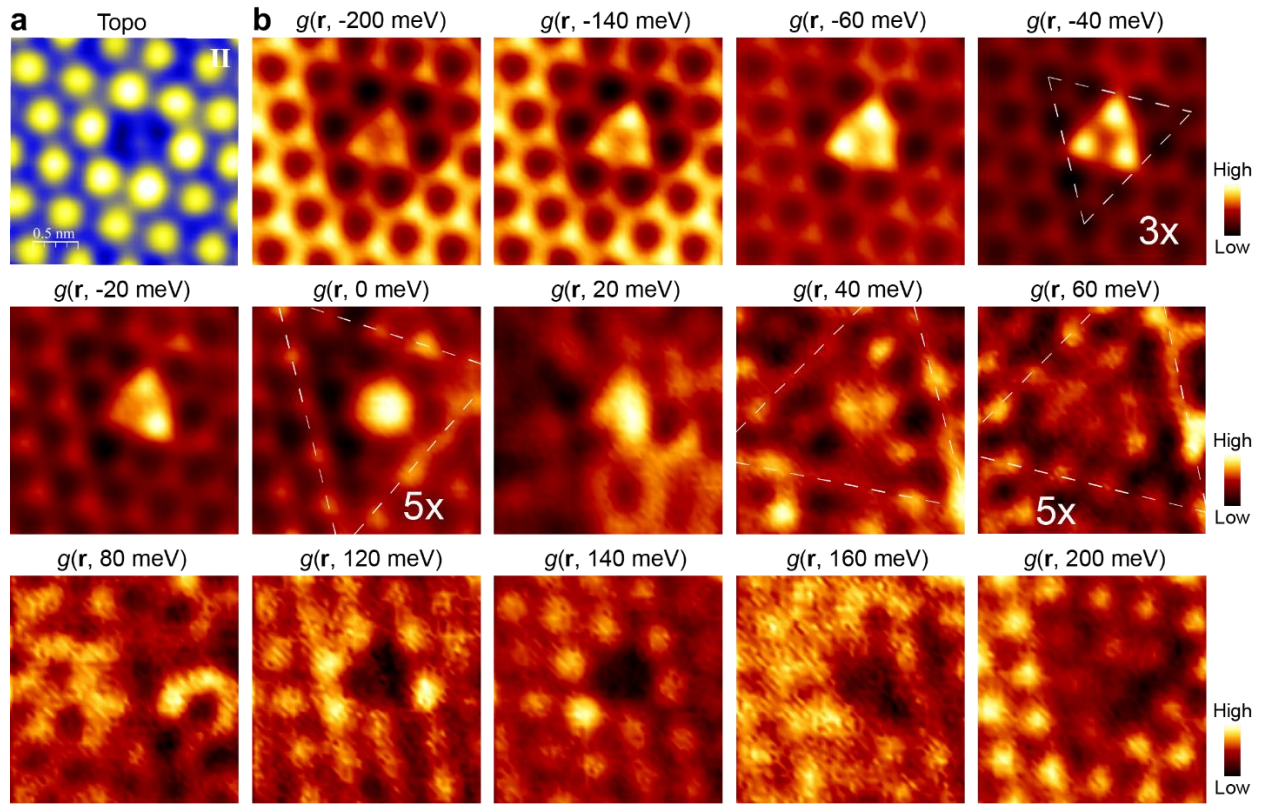

**Figure 2. Anisotropic QPI near a Sn vacancy on the K layer in FeSn/STO(111) films. a**, Topographic STM image of the Sn vacancy (type II defect) on the K layer, setpoint: $V$ = -0.5 V, $I$ = 3.0 nA. **b**, dI/dV maps around the defect, setpoint: $V$ = 0.6 V, $I$ = 5.0 nA, $V_{mod}$ = 6 mV.

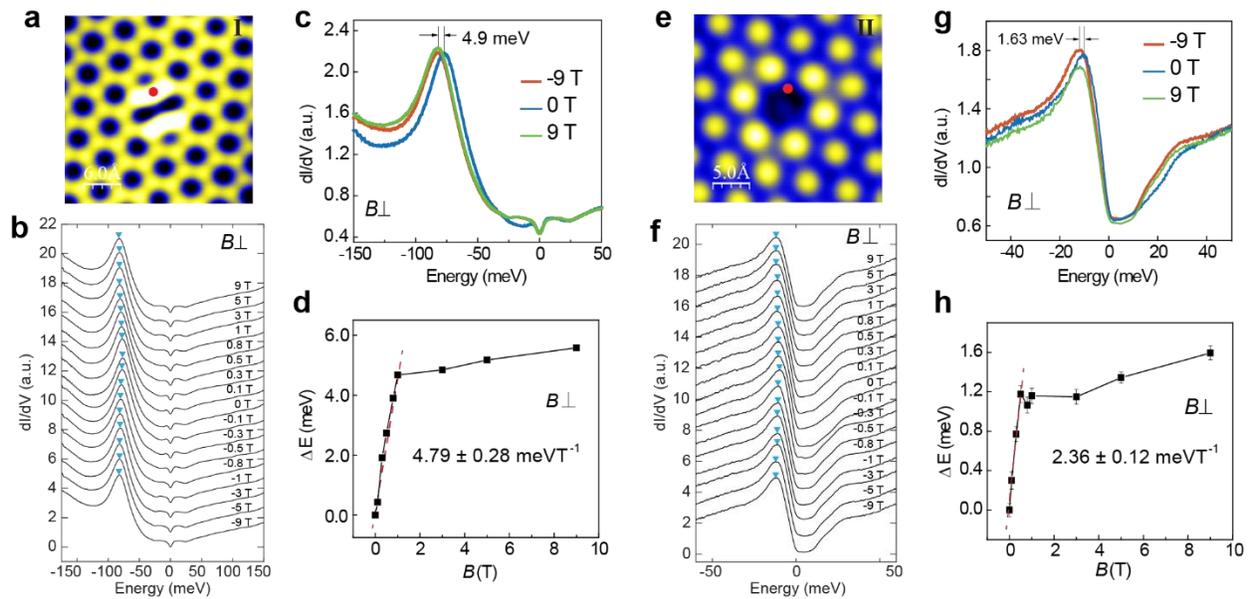

**Figure 3. Anomalous Zeeman shift of the defect bound states on the S and K layers. a**, Topographic STM image of Sn divacancy on the S layer. Setpoint: $V$ = 0.2 V, $I$ = 3.0 nA. **b**, dI/dV spectra taken under vertical magnetic fields as specified. The energy position of the bound states is marked by cyan triangles. **c**, Comparison of the bound states taken at the red dot under 0 T and ± 9 T. A relative shift of 4.9 meV is observed. **d**, Linear fitting of the bound states $\Delta E$ as a function of magnetic field $B_\perp$ within the range [0, 0.5 T]. The slope is 4.79 ± 0.28 meV·T$^{-1}$, corresponding to an effective $g$ factor of 165.2 ± 9.7. **e**, Topographic STM image of a single Sn vacancy on the K layer, setpoint: $V$ = -0.5 V, $I$ = 3.0 nA. **f**, dI/dV spectra of the bound states under various magnetic fields $B_\perp$. **g**, Comparison of the bound states under 0 T and ± 9 T. **h**, Linear fit of the energy shift of the bound states as a function of $B_\perp$ within the range [0, 0.5T]. The slope 2.36 ± 0.12 meV·T$^{-1}$, corresponding to an effective $g$ factor of 81.4 ± 4.2.

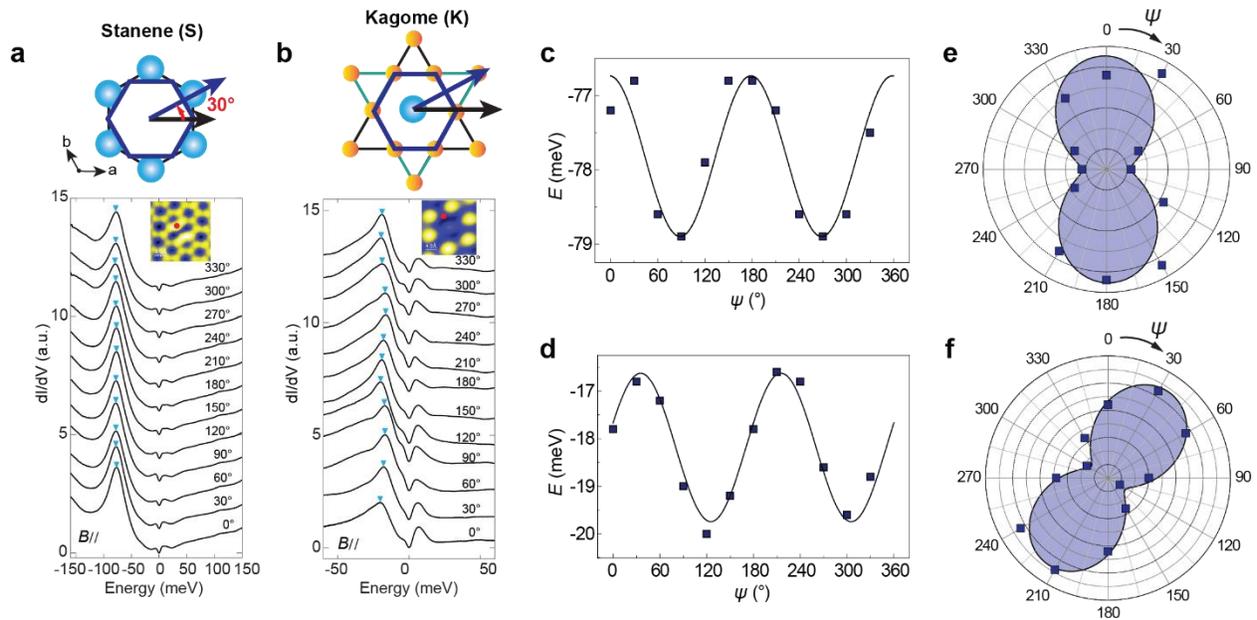

**Figure 4. Two-fold behavior of the defect bound states under in-plane magnetic field. a**, dI/dV spectra under in-plane magnetic fields ($B_{//}$) taken at the red dot of the inset STM image. The magnitude of $B_{//}$ is 1 T, and its direction is denoted by an azimuth angle $\psi$. Inset: STM image of a Sn divacancy on the S layer, setpoint: $V$ = -0.2 V, $I$ = 4.0 nA. **b**, dI/dV spectra under various $B_{//}$ taken at the red dot of the inset STM image. The magnitude of $B_{//}$ is 2 T, and its direction is denoted by an azimuth angle $\psi$. Inset: STM image of a Sn vacancy in the K layer, setpoint: $V$ = 0.2 V, $I$ = 5.0 nA. **c**, The energy positions of the bound state marked by cyan arrows show an oscillatory behavior as a function of $\psi$. The black fitting curve is $E$ = -77.82+1.08*sin(2($\psi$ + 46.53)). **d**, The energy positions of the bound states marked by cyan arrows in (**b**) show an oscillatory behavior as a function of $\psi$. The black fitting curve is $E$ = -18.18+1.56*sin(2($\psi$ + 9.67)). **e-f**, Polar plots of the angle-dependent bound states on the S and K layers. A relative rotation $\psi$ = 29.2° is observed between the S and K layers, consistent with the 30° rotation between the hexagon unit on the S- and K-layer, as schematically shown in the upper panel of **a** & **b**.

**Table of content**

By low-temperature scanning tunneling microscopy/spectroscopy, we report electronic nematicity in Kagome antiferromagnet FeSn films epitaxial on $SrTiO_3$(111) substrate evidenced by the Sn vacancy defect bound states response to vector-magnetic-field.

*Huimin Zhang\*, Zhengfei Wang, Michael Weinert, and Lian Li\**

**Anisotropic response of defect bound states to magnetic field in epitaxial FeSn films**

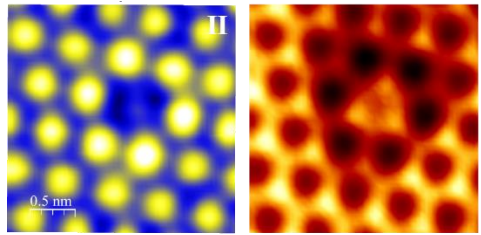